\documentclass[fleqn,twoside]{elsart3}
\usepackage{graphicx}
\usepackage[figuresright]{rotating}

\newcommand{\AmS}{{\protect\the\textfont2
  A\kern-.1667em\lower.5ex\hbox{M}\kern-.125emS}}

\usepackage{epsfig}
\begin{document}
\begin{frontmatter}
\title { Electronic structure of half-metallic magnets } 
\author { B.R.K. Nanda and }
\author { I. Dasgupta\corauthref{cor}}
\corauth[cor]{corresponding author. }
\ead{dasgupta@phy.iitb.ac.in, Tel: +91-22-25767587, FAX: +91-22-25767552}
\address{Department of Physics, Indian Institute of Technology, Bombay, Powai,Mumbai 400 076, India}

\begin{abstract}
We have analyzed the electronic structure 
of half-metallic magnets based on first principles electronic structure 
calculations of a series of 
semi-Heusler alloys. The characteristic feature of the electronic 
structure of semi-Heusler systems is a d-d gap in the 
density of states lying at/close to the Fermi level depending 
on the number of valence electrons. We have employed various 
indicators of chemical bonding to understand the origin of the gap in 
these systems, which is crucial for their half-metallic property. 
The density of states of other half-metallic magnets also supports a gap
and it is a generic feature of these systems. 
We have discussed in some details 
the origin of magnetism, in particular, 
how the presence of the gap is crucial to stabilize 
half-metallic ferro and ferri magnetism in these systems. 
Finally, we have studied the role of magnetic impurities 
in semiconducting semi-Heusler systems. 
We show with the aid of model supercell calculations that these
systems are not only ferromagnetic but also half-metallic with possibly 
high Curie temperature.\\
\begin{keyword}
semi-Heusler \sep  half-metals
\end{keyword}
\end{abstract}
\end{frontmatter}
\section{Introduction}

In the recent times, one of the most growing field of research interest is 
spin-electronics\cite{ref1} (spintronics) where   
the spin of the electron over and above its charge 
is exploited to design new generation of 
electronic devices.
The main ingredient for spintronics
is a source of spin polarized charge carriers. The half-metallic ferromagnets
where one spin direction behaves like a metal and the other is insulating
results in 100\% spin polarization. So during the
spin injection process only electrons of either spin (up or down) 
can be injected into the system
thereby providing an avenue for creation of perfect spin filter
and spin dependent devices.
The half-metallic ferromagnetism was first predicted
by de Groot {\it et. al.}\cite{ref2} by spin-polarized band structure calculations 
in the Mn-based semi Heusler alloy NiMnSb, which is now well established 
experimentally for single crystalline samples.
In addition to Heusler and semi-Heusler 
alloys, the other known 
half-metallic ferromagnetic materials are 
oxides\cite{ref3} (CrO$_{2}$ and Fe$_{3}$O$_{4}$), manganites
La$_{0.7}$Sr$_{0.3}$MnO$_{3}$\cite{ref4}, the double perovskite compound 
Sr$_{2}$FeMoO$_{6}$\cite{ref5}, zinc-blende compounds like CrAs and CrSb\cite{ref6}. 
Half-metallicity can also be induced in an otherwise semiconducting 
or insulating system by incorporation of transition metal impurities
as demonstrated in the diluted magnetic semiconductors\cite{ref7} like 
(In,Mn)As, (Ga,Mn)N, Co substituted TiO$_{2}$ and ZnO \cite{ref8}.
The current research in spintronics is directed 
toward an extensive search for new half-metallic materials with high 
Curie temperature. Also there are considerable effort 
to understand the physics responsible 
for the novel magnetic properties exhibited by these materials as improving 
this understanding, is likely to help in  engineering 
these novel materials. In this communication, we shall address some of 
these issues
by analyzing the electronic and magnetic structure of a series 
of Fe,Co and Ni based half-metallic semi-Heusler alloys.

\begin{figure}[ht]
\begin{center}
\epsfig{figure=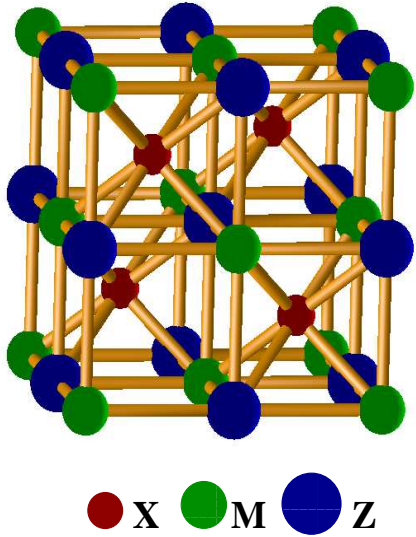,width=0.40\textwidth,height=0.35\textheight}
\end{center}
\caption{Crystal structure of semi-Heusler compounds. }
\end{figure}
The half-metallic semi-Heusler alloys with the general formula XMZ where 
X and M are transition metals and Z is a sp-valent element can be 
particularly attractive for spintronics applications 
due to their relatively high Curie 
temperature and the similarity of their crystal structure 
to the zinc-blende structure adopted by a large 
number of semiconductors like GaAs, ZnSe, InAs {\it etc} 
making them compatible with semiconductor technology.
The key feature of the electronic structure 
of these materials is a gap close to the Fermi level and plays a crucial 
role to drive some of these systems to half-metallic ferromagnets.
The gap in the electronic structure, seems to be a  generic feature of 
half-metallic magnets\cite{pick} resulting from a unique combination of 
their novel crystal structure and chemical bonding. So it is crucial 
to understand the origin of the gap as well as
to identify which factors (structural and chemical)
that influences the gap. For the semi-Heusler systems, 
it has been found that the lattice parameter, 
the relative ordering of the atoms in the unit cell,  
reduced dimensionality(surfaces), chemical substitution(doping) and disorder
have profound influence on the gap and 
hence the half metallic property \cite{dedrev}. 
So a proper coordination of all these factors is crucial for designing these 
materials for possible applications. In this respect, we have investigated 
the role of magnetic impurities in some semiconducting semi-Heusler 
systems. We find even for impurity concentration as low as 
3 $\%$,  some semiconducting semi-Heusler systems 
can be transformed into half-metallic 
ferromagnets  with possibly high Curie temperature. In this 
communication, we shall also discuss
the electronic structure and origin of ferromagnetism in Mn 
doped semiconducting semi-Heusler alloy NiTiSn.\\
\section{Crystal and paramagnetic electronic structure of semi-Heusler systems}
The semi-Heusler compounds XMZ (X=Fe,Co,Ni; M=Ti,V,Cr,Mn,Mo; Z= Sn,Sb)
studied in this work crystallize in the face centered cubic structure
with one formula unit per unit cell as shown in Fig. 1. The space group is
F4/3m (No 216).  
The M and Z atoms are located at 4a(0,0,0) and
4b($\frac{1}{2}$,$\frac{1}{2}$,$\frac{1}{2}$) positions forming the rock
salt structure arrangement. The X atom is located in the octahedral
coordinated pocket, at one of the cube center positions 4c($\frac{1}{4}$,
$\frac{1}{4}$,$\frac{1}{4}$) leaving the other 4d($\frac{3}{4}$,$\frac{3}{4}$,
$\frac{3}{4}$) empty. When the Z-atomic positions are empty
the structure is analogous to zinc blende structure which is common for 
large number of semiconductors. 
All the electronic structure calculations reported in this work
have been done using the experimental lattice constant, except for 
FeMnSb, FeCrSb and  CoMoSb 
where we have estimated theoretically the equilibrium lattice constant
using full potential (FP) LMTO method as described in \cite{ref12}.
The analysis of the electronic structure and chemical bonding 
is carried out in the framework of the 
tight-binding linearized muffin tin orbital method (TB-LMTO) in the 
atomic sphere approximation (ASA) \cite{ref9}
within LDA \cite{ref10} as well as GGA \cite{ref11}.
\begin{figure}[ht]
\begin{center}
\epsfig{figure=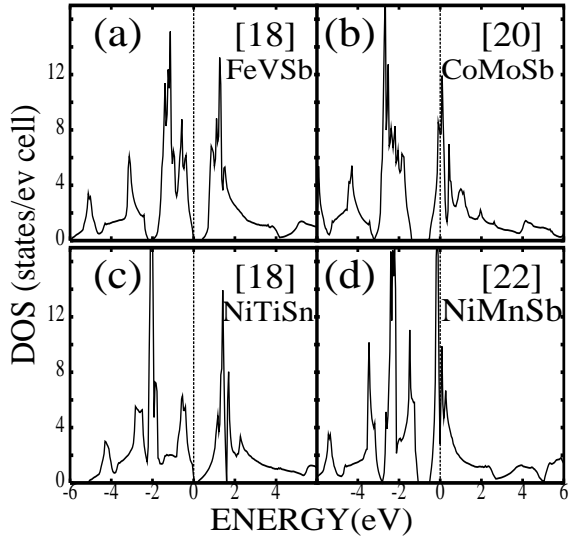,width=0.50\textwidth,height=0.37\textheight}
\end{center}
\caption{Paramagnetic total DOS for semi-Heusler systems (see text).
 The number of valence electron is indicated in the figure. All energies are w.r.t the Fermi energy.}
\end{figure}
In Fig. 2, 
we have displayed the paramagnetic total density of states (DOS)
for the compounds FeVSb, CoMoSb, NiTiSn, and NiMnSb
with the valence electron count (VEC) ranging from 18 to 22
electrons.  From Fig. 2,
we gather that the characteristic feature of all
semi-Heusler compounds
considered here is a gap at/very close to the Fermi level. In addition, 
there is also a virtual gap far below the Fermi level. 
The former is a d-d gap resulting
from the covalent hybridization of the higher valent transition element
X with the lower valent transition element M while the latter
is due to the
X(Fe,Co,Ni)-d-Sb-p interactions  \cite{ref12}, \cite{dedhh}.
As we increase the number
of valence electrons, the bands are progressively filled
so that the 18 electron compounds are narrow gap semiconductors while
the others with less or more than 18 valence electrons
are metals in the paramagnetic phase.
The states
lying below the p-d gap are Sb p-sates. In fact, the Sb-s state is lying
further below the chosen scale of the figure.
The states lying below and above
the d-d gap are the bonding and antibonding
states resulting from the covalent hybridization of X with M.
The bonding bands are predominantly of X-d character while the
antibonding
bands are predominantly M-d character. Hence below the d-d gap
there are 9 bands (4 Sb s+p, 5 predominantly X-d)
which in the  paramagnetic state can accommodate 18 electrons.
With 18 valence electrons all the bonding orbitals are occupied
leading to saturation of otherwise highly unsaturated metallic bonds
providing directionality and strong bonding. As a result,
for the 18 electron compounds the Fermi level 
lies at the edge of the gap and make them semiconductors.
However if there are more than 18 valence electrons the antibonding bands gets
occupied and the paramagnetic state may no longer
be stable. Such instabilities may be
alleviated by the formation of magnetic phase to be discussed in the 
next section.
\\
\\

\begin{figure}[ht]
\begin{center}
\epsfig{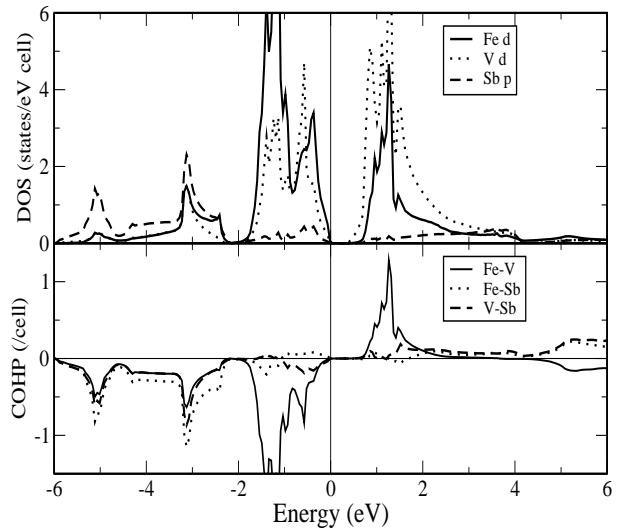}
\end{center}
\caption{Site projected DOS (top) and COHP (bottom) for FeVSb. All energies are w.r.t. Fermi energy.}
\end{figure}
In order to understand the physical origin of the 
gap we have calculated 
the crystal orbital Hamiltonian population (COHP) \cite{cohp}
for various pairs of atoms, as it provides the information concerning the
relative contributions to bonding arising from 
different interactions in the system. In COHP, we calculate the DOS 
weighted by the Hamiltonian matrix elements. The on-site COHP, 
corresponds to the atomic contribution and the off-site COHP covalent 
contribution to the bands. In Fig. 3, we have displayed the 
site projected DOS 
for a representative compound FeVSb 
and the off site COHP for the nearest neighbor Fe-V, Fe-Sb and V-Sb 
interactions. The bonding contribution for which the system 
undergoes a lowering of energy is indicated by negative COHP and the
antibonding contribution that raises the energy is represented by positive COHP.From the COHP plot in Fig. 3 we gather that the most dominant
nearest neighbor interaction is between Fe and V.
A comparison with the partial DOS reveals
that the bonding and the antibonding states below and above the
Fermi level is due to the the nearest neighbor
Fe-V interaction and this is the key interaction
to open up gap close to the Fermi level in these compounds.
The occupied Sb states below the p-d gap are a result of
Fe-Sb and V-Sb bonding interactions with a relatively dominant
contribution from the Fe-Sb interactions.
The presence of Sb atom which provides
a channel to accommodate some transition metal d electrons
in addition to its sp electrons
is therefore crucial
for the stability of these systems. 
\section{Spin polarized Calculations }
\begin{figure}[ht]
\begin{center}
\epsfig{figure=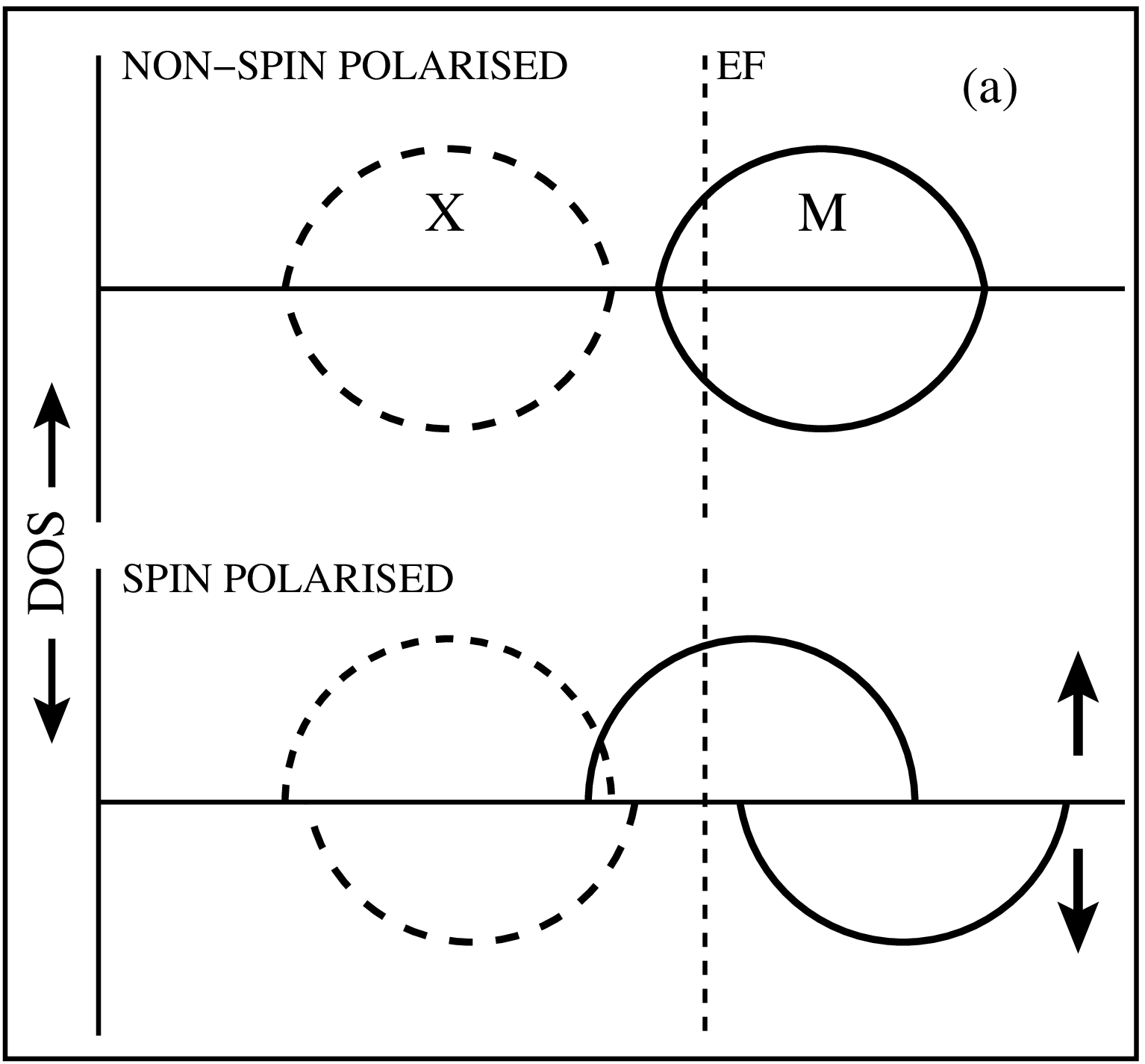,width=0.44\textwidth,height=0.30\textheight}
\epsfig{figure=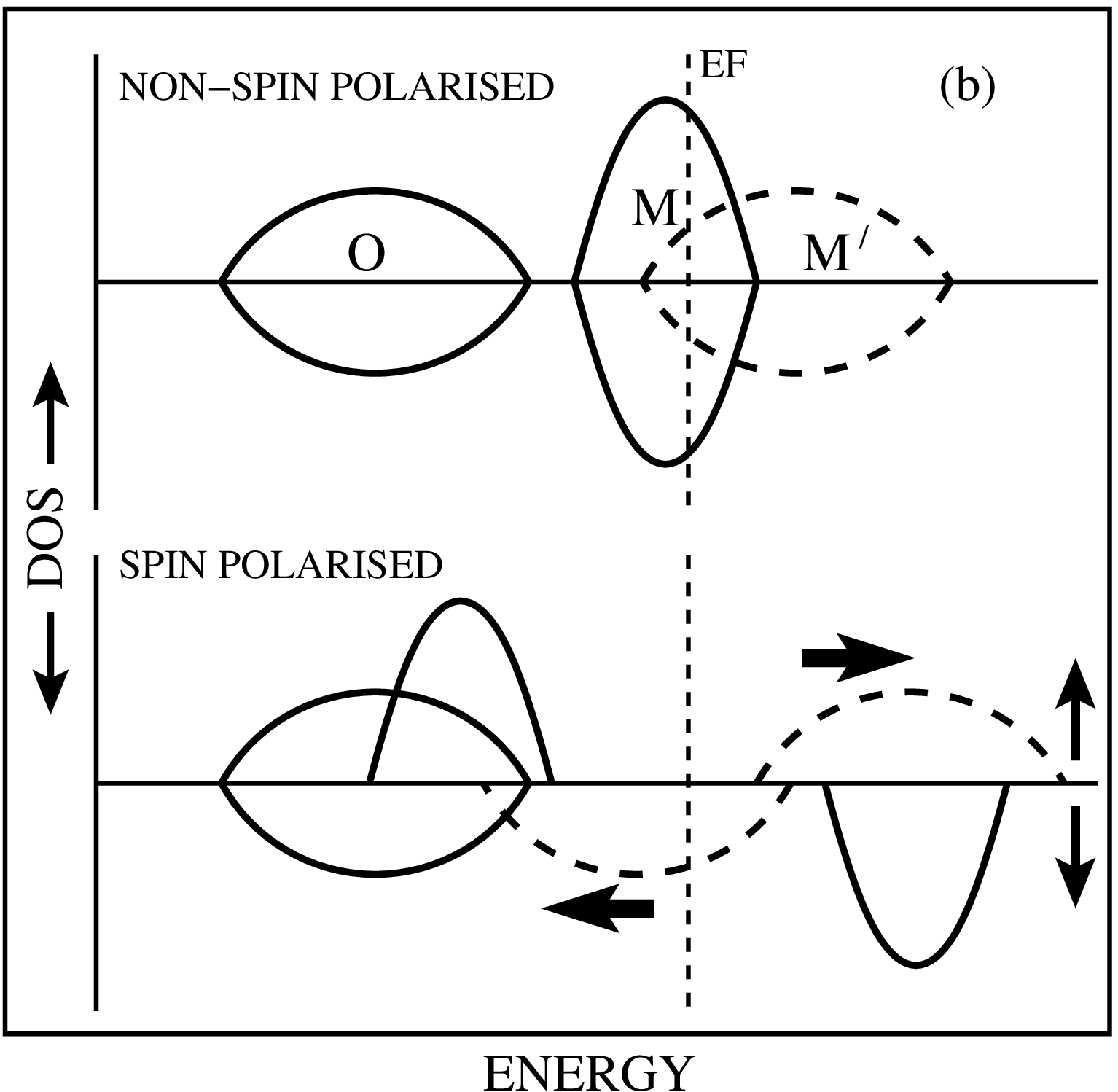,width=0.44\textwidth,height=0.30\textheight}
\end{center}
\caption{Schematic diagram within rigid band model 
explaining the origin of 
half-metallic magnetism (a)  for
semi-Heusler phases(XMZ) and (b) for some double perovskite systems 
$(Sr_{2}MM'O_{6})$ {\it e.g.}
Sr$_{2}$FeMoO$_{6}$}
\end{figure}
\begin{figure}[ht]
\begin{center}
\epsfig{figure=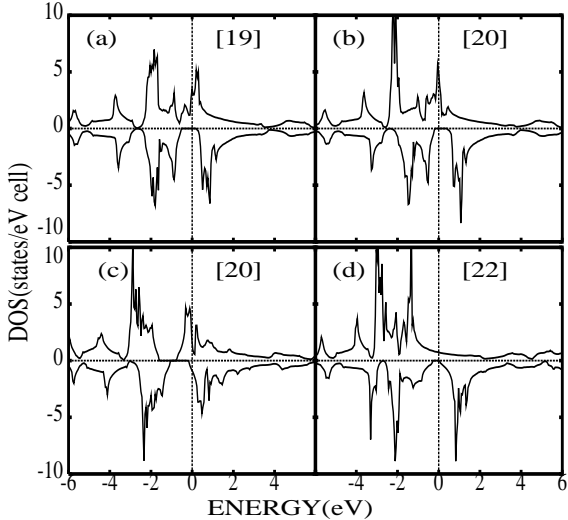,width=0.5\textwidth,height=0.36\textheight}
\end{center}
\caption{Spin-Polarized DOS for (a)FeCrSb, (b)FeMnSb, (c)CoMoSb, (d)NiMnSb. All
 energies are w.r.t. Fermi energy.}
\end{figure}
The characteristic feature of the paramagnetic 
electronic structure of the semi-Heusler compounds discussed in the 
preceding section 
is a d-d gap close to the Fermi level.  The paramagnetic
electronic structure of
other half metallic systems like CrO$_{2}$\cite{ref13}, Fe$_{3}$O$_{4}$\cite{ref14}, 
double perovskite Sr$_{2}$FeMoO$_{6}$\cite{ref15}, zinc blende compounds also sustains a 
gap close to the Fermi level and this seems to be a generic feature 
of half metallic systems. The presence of the gap \cite{ref12}, \cite{theochem}
has an interesting consequence for the 
magnetic properties of these systems. If the Fermi level lies in the 
antibonding complex and/or the density of states 
at the Fermi level is  high then the paramagnetic  
state may no longer be stable. The stability can be achieved 
by developing magnetic order, this is due to the fact
that upon spin polarization, the electrons arrange themselves so that the 
spontaneous magnetization makes the spin-up and spin-down electrons 
different. As a result, the overall bonding energy (gain in kinetic energy
and reduction of Coulomb repulsion due to Hund exchange) is lowered
to make the system stable.
If the paramagnetic electronic structure supports a gap 
then in the process of spin polarization, depending on the 
position of the gap and the Fermi level, the rearrangement of the electrons,
{\it i.e.}
the depletion of the minority bands or the 
occupancy of the majority bands may happen in such way that 
it stabilizes half-metallic ferromagnetism. 
In figure 4(a) we have shown schematically the case relevant for 
semi-Heusler alloys. For semi-Heusler alloys(XMZ) with more than 
18 VEC the Fermi level is at the antibonding complex of predominantly 
M character and the gap is below the Fermi level (figure 4(a), Z states
are not shown).
Now the depletion of the minority antibonding 
states above the gap may happen in such a way 
that none of the
minority antibonding states are occupied
leading to complete spin polarization and we have a half-metallic 
ferromagnet. To stabilize the half-metallic ferromagnetic state
by this  
mechanism the other important requirement is appreciable exchange splitting
of the bands undergoing spin polarization. 
In Fig 5 we display, spin polarized DOS for FeCrSb, FeMnSb, CoMoSb 
and NiMnSb. We gather from the figure that the large exchange splitting 
of Mn and Cr are crucial to stabilize half metallic ferro magnetism,
while CoMoSb is magnetic but not half metallic owing to the weak exchange
splitting of the more extended Mo 4-d bands.

There is however, another possibility, to stabilize half-metallic 
magnetism. Such a mechanism for {\it e.g.}
is realized in double perovskites Sr$_{2}$MM$^{\prime}$O$_{6}$
(M=Fe, M$^{\prime}$=Mo,Re) where 
the paramagnetic DOS is such that 
the states close to the Fermi level have both localized 
states (M) which are 
capable of sustaining large exchange splitting as well as 
delocalized states(M$^{\prime}$)
with weak exchange splitting.  It has 
been recently proposed \cite{ref15}, \cite{kanamori}
that ferromagnetism in these systems can be stabilized by 
hybridization induced negative exchange splitting. 
In Fig. 4(b) we have schematically illustrated this mechanism
relevant for half-metallic double perovskites like 
Sr$_{2}$FeMoO$_{6}$ where in the absence of any spin polarization Fe-d derived 
localized states (M)
as well as Mo-O derived conducting states(M$^{\prime}$)
 are in close vicinity 
of the Fermi level (Fig. 4(b) ). 
In the process of spin polarization, the 
localized M derived states will 
exchange split leaving the weakly exchange split
M$^{\prime}$
derived states at the Fermi level if there is no 
hybridization between M and M$^{\prime}$ states.
However, in the presence of hybridization which is rather strong 
in these systems, there is hybridization induced negative exchange splitting
of M$^{\prime}$ (indicated by the arrows in Fig. 4(b))
leading to antiferromagnetic coupling between the localized M and the 
itinerant M$^{\prime}$ 
 electron states. However, in order to support this 
large gain in energy by antiferromagnetic coupling the localized M 
states should remain ferromagnetically coupled. The presence 
of the gap coupled with the  strong hybridization induced negative 
exchange splitting  drives the system 
to be a half-metallic ferrimagnet.

\section{Electronic structure of doped semi-Heusler systems }
\begin{figure}[ht]
\begin{center}
\epsfig{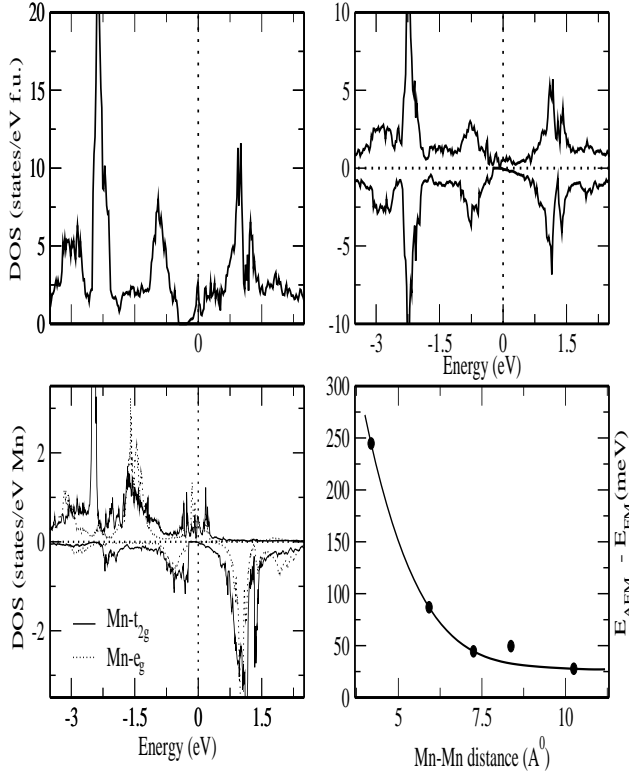}
\end{center}
\caption{ NiTi$_{1-x}$Mn$_{x}$Sn x=6.25$\%$,
(a) paramagnetic total DOS, (b) spin-polarized total DOS, 
(c) Mn-t$_{2g}$ and Mn-e$_{g}$ spinpolarised DOS, 
(d)Energy difference between aniferromagnetic 
and ferromagnetic states for various Mn-Mn distances in the supercell.}
\end{figure}
In this section, we shall discuss the role of magnetic impurity
Mn in the semiconducting semi-Heusler alloy NiTiSn having 18 valence 
electrons. 
In order to study the effect of the
dilute Mn impurity in semi-Heusler systems ({\it i.e. } to
simulate the effect of doping) we have constructed supercells with
size dependent on the $\%$ of doping. In each supercell we have replaced
two Ti  atoms by two Mn atoms which also allowed us to study
the interaction between Mn moments.  The size of the supercell
was chosen to ensure that the separation between the impurities
is much smaller in comparison to the dimension of the supercell.
Replacing a Ti atom with Mn atom puts three additional d-electron to the system
per Mn atom, {\i.e.} $\Delta Z$ = 3.
As a result, VEC $>$ 18 and the system is not 
semiconducting and according to our previous discussion 
it is likely to stabilize in the 
ferromagnetic state. In Fig. 6(a) we show the total paramagnetic DOS for 
NiTi$_{1-x}$Mn$_{x}$Sn (x=6.25 $\%$) 
system. The crucial feature of the DOS is 
the presence of a shallow donor level produced by the addition of Mn impurities
and the Fermi level lies on this Mn impurity driven state. In order to check
whether the system sustains magnetic instability we have performed 
spin polarized density functional calculations in the framework 
of GGA with (i) two Mn spins parallel to each other, 
the ferromagnetic (FM) configuration and (ii) 
two Mn spins antiparallel to each other the 
antiferromagnetic (AFM) configuration. Our calculations within 
LMTO-ASA method shows that the ferromagnetic state is most stable 
for all concentrations, ranging from 25$\%$ Mn to 3.125$\%$ Mn. 
The doped systems are not only ferromagnetic but also half metallic 
sustaining a magnetic moment of 3 $\mu_{B}$ per Mn for all concentration ranges.
In figure 6(b) we have shown the spin polarized DOS for 
the same system, indicating that the system is half-metallic.
Here the defect states are screened metallically by the majority states
so that the number of the minority states do not change and in this case  the alloy moment is given by $m = \Delta Z$.
The analysis of the DOS close to the Fermi level suggests that it is
predominantly of Mn character. The Mn atoms substituting Ti 
are in tetrahedral arrangement 
with Ni,  as a consequence Ni $t_{2g}$ - Mn $t_{2g}$ interactions are more favorable.
The antibonding states near the Fermi level are of 
Mn t$_{2g}$ character, while the bonding states are Ni t$_{2g}$ like. 
However a bonding partner of Mn t$_{2g}$ like state is seen as a sharp peak 
at about -2 eV below the Fermi level. The e$_g$ like states are nonbonding.
So out of the additional three electrons available, two are accommodated 
in the e$_g$ like sates keeping the t$_{2g}$ like state partially empty. 
This can be seen in Fig. 6(c). Such 
partially filled t$_{2g}$ states are favorable for ferromagnetism
as the 3d electrons in the partially occupied 3d-orbitals is allowed to hop 
to the neighboring 3-d states provided the spin are parallel. 
The lowering of energy by hopping in the presence of parallel alignment of spin the so called double exchange mechanism stabilizes ferromagnetism.
However if the band were empty such a lowering of energy is not possible 
and the AFM arrangement of impurity spin will be favored. Finally in Fig. 
6(d) we show the energy difference $\Delta E = E_{AFM}-E_{FM}$ as a function of 
separation of the impurity. We see from the figure that $\Delta E$ 
decreases sharply with distance suggesting that the double exchange induced 
ferromagnetic interaction is short ranged. This indicates the formation of 
Mn clusters within a short radial distance and might lead to high values of 
$T_{c}$ in these systems. Based on this theoretical studies it will be interesting to investigate these systems experimentally.
\ack{We thank Dr. S.B. Roy and Dr. T. Saha-Dasgupta for useful 
discussions. Financial support from FDF IIT-B and INSA 
to attend ACCMS-2 
is gratefully acknowledged.
 BRKN thanks CSIR, India, for research fellowship (SRF). The research is funded by CSIR (grant no. 03(0931)/01/EMR-II).}

\end{document}